\documentstyle[prc,12pt,aps]{revtex}

\def\diag{\mathop{\rm diag}\nolimits}
\def\bm#1{\mbox{\boldmath$#1$}}

\begin{document}


\title{
Consistent BCS+RQRPA formalism with application to 
the double beta decay}


\author{A. Bobyk, Wies{\l}aw A. Kami{\'n}ski, P. Zar{\c e}ba}
\address{Department of Theoretical Physics, Maria Curie--Sk{\l}odowska 
University,\\
Radziszewskiego 10, 20--031 Lublin, Poland}

\date{\today}

\maketitle


\begin{abstract}

A new consistent analysis of the renormalized proton--neutron 
quasiparticle random phase approximation
based on the simultaneous recalculation of the one--body 
density matrix and the pairing tensor has been used to
study the double beta decay. 
We demonstrated that inclusion of the quasiparticle correlations 
at the BCS level reduces the ground state correlations in the
particle--particle channel of the proton--neutron interaction.
We also simplified the RQRPA equations significantly
obtaining a low--dimensioned set of linear equations for
the quasiparticle densities.
The formalism was applied to the double beta decay
in $^{76}$Ge. 

\end{abstract}
   





The proton--neutron quasiparticle random phase approximation (pn--QRPA)
has been considered the most powerful method for the beta and double beta
transition calculations of nuclear sytems which are far away from 
the closed shells \cite{Vog86,Civ87,Tom87,Eng88,Mut88,Suh88,Mut89,Pan92}.
A remarkable success was achieved especially by the QRPA approach
in revealing the suppression mechanism of the neutrino accompanied
double beta decay, a long--standing problem of the theoretical
treatment of this process.
Further development of the approach went beyond many shortcomings
and refined calculations of nuclear matrix elements involved 
in the double beta decay.  
Among others, the following problems were set and solved: 
particle number nonconservation \cite{Civ91,Krm93}, 
role of the proton--neutron pairing \cite{Che93}, violation of 
the Pauli exclusion principle \cite{Tsuh95,Sch96}, 
higher--order corrections to the ordinary QRPA \cite{Rad91,Sto93}, 
treatment of transitions to final excited states 
\cite{Suh93,Bob95,Tsuh97,Schw98},
extension of the definition of phonon operators by means of
the--so--called scattering terms \cite{Rad98}, etc. 

Most of such improvements disregarded, however, the main source of the
formalism instability connected with violation of the Pauli 
exclusion principle by using the commutation relations for the QRPA phonon
operators. 
To overcome this shortcoming of the pn--QRPA framework the renormalization
technique was proposed \cite{Tsuh95} and extended to include 
the proton--neutron pairing \cite{Sch96}.
The main goal of the method called in literature the renormalized
QRPA (RQRPA) is to use a self--iteration of the 
QRPA equation to take into account the additional one--quasiparticle
scattering terms in the commutation relations.
But this procedure results in a non--vanishing quasiparticle content
of the ground states and followed by some inconsistency between
RQRPA and the BCS approach since the ground state approximated 
by the BSC state is chosen to be the quasipatricle vacuum.  
To minimize the influence of such a discrepancy one needs
to reformulate the BCS equations in a way proposed in \cite{gsc}.
Combining both RQRPA and so modified BCS one obtains the 
self--consistent BCS+RQRPA approach (SRQRPA) which we study in more detail 
in this paper.





In the QRPA (either ordinary or renormalized) approach one assumes the
harmonicity of the nuclear motion and starts with the excited--state
creation phonon operators of the form \cite{rowe,baranger}:
\begin{equation}
  Q^{m\dagger}_{J^\pi M} = 
  \sum_{pn} \left[ X^m_{(pn)J^\pi} A^\dagger_{(pn)J^\pi M}
  - Y^m_{(pn)J^\pi} \tilde{A}_{(pn)J^\pi M} \right],
\label{Qdagger}
\end{equation}
where $X^m_{(pn)J^\pi}$ and $Y^m_{(pn)J^\pi}$ are the
forward--going and backward--going variational amplitudes, respectively. 
The quantities $A^\dagger_{(pn)J^\pi M}\equiv[a_p^\dagger 
a_n^\dagger]_{J^\pi M}$ are the angular--momentum coupled 
two--quasiparticle creation operators. 
Since they do not fulfil the bosonic commutation relations exactly
in the quasiboson approximation (QBA), that is used to derive the usual
QRPA equations, the Pauli principle is violated. 
To avoid this serious drawback in the improved version of the theory 
one introduces the renormalized operators \cite{catara}:
\begin{equation}
  {\cal A}^\dagger_{(pn)J^\pi M} \equiv D_{pn}^{-1/2} A^\dagger_{(pn)J^\pi M},
\end{equation}
along with the renormalized amplitudes:
\begin{eqnarray}
  {\cal X}^m_{(pn)J^\pi} & \equiv & D_{pn}^{1/2} X^m_{(pn)J^\pi}, \\
  {\cal Y}^m_{(pn)J^\pi} & \equiv & D_{pn}^{1/2} Y^m_{(pn)J^\pi},
\end{eqnarray}
where $D_{pn}$ matrix is defined by the expectation value in the RPA
ground--state of the commutator:
\begin{equation}
D_{pn} \equiv \left<0\left|[A_{(pn)J^\pi M},A^\dagger_{(pn)J^\pi
M}]\right|0\right>   = (1 - n_p - n_n)
\label{D_pn}
\end{equation}
together with the quasiparticle densities
($\hat{\jmath}_a\equiv\sqrt{2j_a+1}$):
\begin{eqnarray}
n_p & \equiv & \hat{\jmath}_p^{-1} \left<0\left|
[a_p^\dagger\tilde{a}_p]_{00}\right|0\right>, \\ 
n_n & \equiv & \hat{\jmath}_n^{-1}\left<0\left|
[a_n^\dagger\tilde{a}_n]_{00}\right|0\right>.  
\end{eqnarray}
The above equation has been derived using the exact fermionic commutation
relations and thus it goes beyond the ordinary quasiboson approximation.
One can prove easily that now the following relation holds:
\begin{equation}
\left<0\left|[{\cal A}_{(pn)J^\pi M}, {\cal
A}^\dagger_{(p'n')J'^{\pi'}M'}]\right|0\right> = 
  \delta_{pp'}\delta_{nn'}\delta_{JJ'}\delta_{\pi\pi'}\delta_{MM'},
\end{equation}
i.e.\ the renormalized operators behave as bosons, at least in the sense
of the ground--state expectation value of their commutator. 
The phonon operator now reads:
$$
  Q^{m\dagger}_{J^\pi M} = 
  \sum_{pn} \left[ {\cal X}^m_{(pn)J^\pi} {\cal A}^\dagger_{(pn)J^\pi M} 
  - {\cal Y}^m_{(pn)J^\pi} \tilde{\cal A}_{(pn)J^\pi M}\right],
\eqno{(\ref{Qdagger}')}
$$
and using e.g.\ the equation of motion (EOM) method \cite{rowe} one gets
the RQRPA equations of the usual form:
\begin{equation}
  \left(
    \begin{array}{cc}
       {\cal A} & {\cal B} \\ 
       {\cal B} & {\cal A}
     \end{array}
   \right)_{J^\pi}\left(
     \begin{array}{c}
       {\cal X}^m \\
       {\cal Y}^m
     \end{array}
   \right)_{J^\pi} = \Omega^m_{J^\pi}
   \left(
     \begin{array}{c}
       {\cal X}^m \\
      -{\cal Y}^m
     \end{array}
   \right)_{J^\pi},
\label{RQRPA}
\end{equation}
with the new renormalized RPA matrices ${\cal A}$ and ${\cal B}$ defined
in \cite{Tsuh95}. 
Here $\Omega^m_{J^\pi}\equiv E^m_{J^\pi}-E_0$ is the RPA excitation 
energy with respect to the ground--state.

Now there appears the question of the calculation of the $D_{pn}$ matrix 
entering the expressions for the RPA matrices ${\cal A}$ and ${\cal B}$. 
Using the mapping \cite{Tsuh95}:
\begin{eqnarray}
  [a_p^\dagger\tilde{a}_p]_{00} & \mapsto & \hat{\jmath}_p^{-1}
  \sum_{J^\pi Mn} A^\dagger_{(pn)J^\pi M} A_{(pn)J^\pi M}, {}
\label{mapping1}
\end{eqnarray}
\begin{eqnarray}
  [a_n^\dagger\tilde{a}_n]_{00} & \mapsto & \hat{\jmath}_n^{-1}
  \sum_{J^\pi Mp} A^\dagger_{(pn)J^\pi M} A_{(pn)J^\pi M},
\label{mapping2}
\end{eqnarray}
and inverting (\ref{Qdagger}) one derives the following equations for the
quasiparticle densities:
\begin{eqnarray}
  n_p & = & \hat{\jmath}_p^{-2} \hat{J}^2 \sum_{J^\pi mn} D_{pn}
  |{\cal Y}^m_{(pn)J^\pi}|^2, \\
  n_n & = & \hat{\jmath}_n^{-2} \hat{J}^2 \sum_{J^\pi mp} D_{pn}
  |{\cal Y}^m_{(pn)J^\pi}|^2.
\label{n_p and n_n}
\end{eqnarray}
Insterting (\ref{D_pn}) into (\ref{n_p and n_n}) one gets the following
system of linear equations for $n_p$ and $n_n$:
\begin{eqnarray}
  {\cal Y}'_p n_p + \sum_n {\cal Y}_{pn} n_n & = & {\cal Y}_p, \nonumber  \\
  {\cal Y}'_n n_n + \sum_p {\cal Y}_{pn} n_p & = & {\cal Y}_n,
\label{lineq_n}
\end{eqnarray}
or, in the matrix form:
$$
  \left(\begin{array}{cc}
    \diag({\cal Y}_{(p)}') & {\cal Y} \\
    {\cal Y}^{\rm T} & \diag({\cal Y}_{(n)}')
  \end{array}\right)
  \left(\begin{array}{c}
    n_{(p)} \\ n_{(n)}
  \end{array}\right) =
  \left(\begin{array}{c}
    {\cal Y}_{(p)} \\ {\cal Y}_{(n)}
  \end{array}\right), \eqno{(\theequation')}
$$
where
\begin{eqnarray}
  {\cal Y}_{pn} & \equiv & \hat{J}^2 \sum_{J^\pi m} |
  {\cal Y}^m_{(pn)J^\pi}|^2,
  \nonumber \\
  {\cal Y}_p & \equiv & \sum_n {\cal Y}_{pn}, 
  \quad {\cal Y}'_p \equiv \hat{\jmath}_p^{2} + {\cal Y}_p, \\
  {\cal Y}_n & \equiv & \sum_p {\cal Y}_{pn}, 
  \quad {\cal Y}'_n \equiv \hat{\jmath}_n^{2} + {\cal Y}_n. \nonumber
\end{eqnarray}
It is worth mentioning, that the dimension of our linear problem is only
$2n\times 2n$, where $n$ is the dimension of the single-particle basis.
This is of much advantage, since (\ref{lineq_n}) has to be solved many
times as one should iterate between (\ref{RQRPA}) and (\ref{lineq_n})
until convergence is achieved. 
The other way round, inserting (\ref{n_p and n_n}) into (\ref{D_pn}), 
as it has been  done by several authors, e.g.\
\cite{Tsuh95,Sch96}, one obtains the equation:
\begin{equation}
  D_{pn} = 1 - \hat{\jmath}_p^{-2} \sum_{n'} D_{pn'}{\cal Y}_{pn'}
             - \hat{\jmath}_n^{-2} \sum_{p'} D_{p'n}{\cal Y}_{p'n},
\end{equation}
that, on the contrary to the claim expressed in \cite{Tsuh95},
can be transformed into  the $n^2\times n^2$ linear system:
\begin{equation}
  \sum_{p'n'} W_{pn,p'n'} D_{p'n'} = U,
\end{equation}
where
\begin{equation}
  W_{pn,p'n'} \equiv \delta_{pp'}\delta_{nn'} 
    + \delta_{pp'}\hat{\jmath}_p^{-2}{\cal Y}_{pn'}
    + \delta_{nn'}\hat{\jmath}_n^{-2}{\cal Y}_{p'n},
\end{equation}
and $U$ is the vector of 1's. 
In practice however, it takes much less time to solve this system 
using the standard linear algebra procedures than following iteration 
methods, as in \cite{Tsuh95,Sch96}.
But further reduction of the complexity of the problem to the 
form of (\ref{lineq_n}) allows us to take all possible 
multipolarities into account for the calculation of
the $D_{pn}$ renormalization factors. 
Although some of them are less important and neglected 
in \cite{Tsuh95,Sch96}, when one leaves only a few, the J--coupling 
scheme breaks, since the basis becomes incomplete and the validity of 
the mapping (\ref{mapping1})--(\ref{mapping2}) is questionable. 
As can be seen further the results are the evidence for it.

With a non--vanishing quasiparticle content of the ground state one arrives
at the inconsistency between RQRPA and BCS, since in the latter 
assumes the ground state to be the quasiparticle vacuum. 
One thus needs to reformulate the BCS equations \cite{gsc}, namely by
recalculating the density matrix $\rho$ and the pairing tensor $\kappa$.
With the standard Bogoliubov--Valatin transformation \cite{baranger} they
read now:
\begin{eqnarray}
  \rho_a \equiv \left<0\left| c^\dagger_\alpha c_\alpha \right|0\right> 
    & = & v_a^2 + (u_a^2-v_a^2) n_a , \\
  \kappa_a \equiv \left<0\left| \tilde{c}_\alpha c_\alpha \right|0\right>
    & = & u_a v_a (1-2n_a),
\end{eqnarray}
and depend on the quasiparticle densities $n_a$, where $a$ runs over
proton or neutron indices. 
The $u$ and $v$ coefficients are obtained by minimizing the 
ground--state energy, that by virtue of the Wick's
theorem is expressed as \cite{baranger,rs}:
\begin{equation}
  \left<0\left| \hat{H} \right|0\right> = \sum_a \hat{\jmath}_a^2 \varepsilon_a \rho_a
  + \frac{1}{4} \sum_{ab} \hat{\jmath}_a \hat{\jmath}_b
  \left\langle(aa)_{J^\pi=0^+}^{T=1}|V|(bb)_{J^\pi=0^+}^{T=1}\right\rangle 
  \kappa_a \kappa_b
\end{equation}
with the particle--number constraint:
\begin{equation}
N_0 = \left<0\left| \hat{N} \right|0\right> = \sum_a \hat{\jmath}_a^2
\rho_a. 
\end{equation}
In the above, $\varepsilon_a$ are the single--particle energies and
$\left\langle(aa)_{J^\pi=0^+}^{T=1}|V|(bb)_{J^\pi=0^+}^{T=1}\right\rangle$
are the matrix elements of the two--body interaction. 

To solve the SRQRPA equations we start with the ordinary BCS equations,
putting $n_p=n_n=0$, than to proceed with the corresponding RQRPA 
problem (inner iteration), that gives us new quasiparticle densities 
and loop with them back to BCS until 
the convergence is achieved (outer iteration). 
We arrived thus at the doubly--iterative problem and the question 
of efficient and accurate getting through all the calculation
steps becomes very impotrant. 
We then stress again, that without showing that the problem 
of calculating the $D$--matrix can be reduced to the linear system 
(\ref{lineq_n}) of acceptable size the realization of this task would 
be hardly possible.

The ground--state to ground--state $2\nu\beta\beta$ Gamow--Teller matrix
elements are expressed as follows:
\begin{equation}
  M_{\rm GT}^{2\nu} = \sum_{mm'} \frac{
  \langle 0^+_{\rm f,gs}(A,Z+2) || \bm{\sigma}\tau_+ || 1^+_{m'}\rangle
  \langle 1^+_{m'} | 1^+_m\rangle
  \langle 1^+_m || \bm{\sigma}\tau_+ || 0^+_{\rm i,gs}(A,Z) \rangle}
  {\frac{1}{2}\left[\Omega_{1^+}^m+\Omega_{1^+}^{m'}
               +Q_{\beta^-}(A,Z+1)-Q_{\beta^-}(A,Z)\right]},
\end{equation}
where the charge--changing transition densities are:
\begin{eqnarray}
   \langle 0^+_{\rm f,gs} || \bm{\sigma}\tau_+ || 1^+_{m'}\rangle
   & = & \sum_{pn} \langle p || \sigma || n \rangle
     (v'_p u'_n {\cal X}^{m'}_{(pn)1^+} + u'_p v'_n {\cal Y}^{m'}_{(pn)1^+})
     \sqrt{D'_{pn}},
   \\
   \langle 1^+_{m} || \bm{\sigma}\tau_+ || 0^+_{\rm i,gs}\rangle
   & = &\sum_{pn} \langle p || \sigma || n \rangle
     (u_p v_n {\cal X}^m_{(pn)1^+} + v_p u_n {\cal Y}^m_{(pn)1^+})
     \sqrt{D_{pn}}
\end{eqnarray}
and the overlap of intermediate excited states is assumed to be
expressed as:
\begin{equation}
  \langle 1^+_{m'} | 1^+_m\rangle = \sum_{pn}
  \left({\cal X}^{m'}_{(pn)1^+}{\cal X}^m_{(pn)1^+}-
  {\cal Y}^{m'}_{(pn)1^+}{\cal Y}^m_{(pn)1^+}\right).
\end{equation}
In the above the non--primed (primed) quantities result from the SRQRPA
calculations based on the initial (final) ground--state.


To illustrate the differences between the QRPA, the RQRPA and the
SRQRPA and to show much better stability of the SRQRPA solutions, 
 the results of our calculations as a function of
the particle--particle ($g_{\rm pp}$) and particle--hole ($g_{\rm ph}$)
factors, renormalizing the bare two--body NN interaction are plotted
in Figs.\ 1--3. 
This commonly used renormalization is necessary due to tne nucleus
finite size (the bare NN matrix elements are calculated for the 
infinite nuclear matter) and due to the limited dimension of the 
single--particle basis.
In our calaculations we used the two--body matrix elements calculated
from the Bonn--B nucleon--nucleon one boson exchange potential. 
The single--particle energies are calculated from the Coulomb--corrected
Woods--Saxon potential with the Bertsch parametrization. 
We used several sets of single--particle levels to see how the choice of
the basis influences the results.  
We find weak dependence of the RQRPA and the SQRPA results on the 
dimension of the single--particle basis. 
On the other hand, the QRPA shows no stability on the chosen basis. 
The conclusion is that the most suitable basis for the calculations of
$2\nu\beta\beta$ decay in $^{76}$Ge consists of 16 levels with
$^{16}$O as a core. 
Therefore we used this basis in all the further studies described
below. 
To compare with the experiment we have adopted the experimental 
half--life of $T_{1/2}^{2\nu}=(1.42 \pm 0.03 \pm 0.13) \times 
10^{21}\,$yr from the latest measurement by the Heidelberg--Moscow 
$\beta\beta$ cooperation \cite{balysh}. 

In Fig.~1 the calculated double Gamow--Teller matrix elements
$M^{2\nu}_{\rm GT}$ as a function of $g_{\rm pp}$ for two different
$g_{\rm ph}$ values and for three different QRPA approaches are plotted. 
We would like to stress that in these calculations in the self--consistent
iterations of the RQRPA and the SRQRPA all the intermediate
multipolarities were present. 
The comparison between the QRPA, the RQRPA and the SRQRPA results in 
the physically acceptable region of the $g_{\rm pp}$ parameter 
$0.8\leq g_{\rm pp} \leq 1.2$ shows two main features of the 
renormalized QRPA. 
First, the inclusion of the ground--state correlations beyond QRPA 
does not only improve the agreement between theoretical calculations 
and experimental data but also causes the stabilization of the 
dependence of $M^{2\nu}_{\rm GT}$ as a function of $g_{\rm pp}$. 
Second, the iteration procedure for quasiparticle densities which
causes treating of RQRPA and BCS on the same footing stabilizes 
the results even further.

Some authors claim that only the limited set of these multipolarities
plays a role in the evaluation of the double Gamow--Teller matrix elements
\cite{Tsuh95,Sch96}. 
In Fig.~2 there are shown the results of calculations of 
$M^{2\nu}_{\rm GT}$ as a function of $g_{\rm pp}$ for a different number 
of multipolarities for the RQRPA and the SRQRPA, respectively. 
The basis is the same as in Fig. 1, but the value of $g_{\rm ph}$ 
parameter is fixed to 1.0. 
It can be seen why the inclusion of all multipolarities is essential 
to obtain the reliable predictions of the RQRPA and 
the SRQRPA calculations. 
The solid line in Fig. 2 represents the QRPA calculations, 
the dot--dashed line the calculations with only $1^+$ multipolarity, 
the dotted line with multipolarities up to $2^+$, the dashed line up 
to $3^-$, the long--dashed up to $5^-$ and the thick solid line all 
considered multipolarities up to $11^+$.
The inclusion of higher multipolarities causes the shift of the collapse
of the RQRPA and the SRQRPA beyond the value of $g_{\rm pp}=1.0$. 
The additional advantage of the SRQRPA solutions is that the calculated
matrix elements are less dependent on the $g_{\rm pp}$ parameter.

In Fig.~3 the effect of including more multipolarities $J^{\pi}$ in the
RQRPA and the SQRPA calculations of $M^{2\nu}_{GT}$ for fixed $g_{\rm
ph}=1.0$ is shown. 
The filled symbols represent the calculations for $g_{\rm pp}=0.8$ and open
symbols for $g_{pp}=1.0$. 
One can see the saturation effect for higher multipolarities. 
The explanation of this behaviour is that the higher multipolarities 
$J^{\pi}$ are less collective.
Their contribution to the ground--state correlations is much smaller
than the lower ones. 
An interesting feature is the virtual independence of the matrix 
element on the number of multipolarities in SRQRPA around
$g_{\rm pp}=1.0$. 
It is reflected in Fig. 2, where one can see that RQRPA "diverges" when
going with $g_{\rm pp}$ from 0 to 1, while  SRQRPA "converges". 
In our opinion, this is a clear evidence that the self--consistency 
between BCS and RPA, i.e.\ between the ground--state and excited--state 
properties is being restored in SRQRPA.

Finally, we would like to address the question of the Ikeda sum--rule
violation. 
It is well known, that in the usual QRPA the Ikeda sum rule
is conserved exactly if all the spin--orbit partners of the
single--particle orbitals are present in the basis, i.e. 
\begin{equation}
 S_- - S_+ \equiv \sum_m \left| \langle 0^+_{\rm gs} || \bm{\sigma}\tau_+ ||
1^+_m\rangle \right|^2 - \sum_m \left| \langle 0^+_{\rm gs} ||
\bm{\sigma}\tau_- || 1^+_m\rangle \right|^2 = 3(N-Z). 
\end{equation}
 The violation is marginal even if one or two of these partners are 
omitted. 
But this is not the case in RQRPA, where the Ikeda sum rule is violated
up to 20\%\ and is similar, up to 25\%\ in the SRQRPA. 
There is some difference between two nuclei under consideration, i.e. 
germanium, where the situation gets worse when the ground--state 
correlations (SRQRPA) are taken into account, and selenium, where 
the results are slightly improved. 
One can conclude, that there is more to this than the ground--state 
correlations to restore the Ikeda sum rule. 
As already pointed out \cite{ikeda}, the scattering terms present 
in the $\beta$--decay operators can be responsible for this effect
\cite{Krm98}. 
They give no contribution to QRPA, because
they are of the quasiparticle--quasihole character, but they should be 
taken into account when the ground--state quasiparticle densities are not
assumed to be zero, like in the RQRPA or SRQRPA. 
It is necessary to extend the form of the QRPA phonon operator 
(\ref{Qdagger}) by including new excitation modes, the so--called 
$B$--modes \cite{Rad98}.




In conclusion, we have developed a new method of the nuclear 
matrix--element calculations for the neutrino accompanied double beta 
decay to the ground state in a frame of the consistent BCS+RQRPA approach.
Using the $2\nu \beta \beta $--decay in germanium $^{76}$Ge as an example
we demostrated that the inclusion of the ground--state correlations
beyond QRPA causes the stabilization of the dependence of the
Gamow--Teller nuclear matrix element, but also weakens their 
influence on their magnitude because of the additional 
change of the quasiparticle densities during the iteration procedure 
with the modified BCS solution.

Unlike ortodox QRPA which requires fine tuning of the $g_{pp}$ parameter 
describing the particle--particle interaction--strength as the 
matrix element collapses near the physical strength, RQRPA 
and BCS+RQRPA  give stable matrix elements over the whole 
range of physical strength and thus the latter approaches allow
for more predictive power than the old method. 

Owing to the development of the method to calculate the $D_{pn}$
normalization factors we could take all possible multipolarities
into account.
Then we were able to avoid possibile J--scheme breaking because,
by neglecting some of the multipolarities, the basis becomes incomplete
and the validity of mappings (\ref{mapping1})--(\ref{mapping2}) 
is questionable. 



This work has been supported in parts by the State Committee for
Scientific Researches (Poland) and in the frame of PECO 
project, Contract No. ERBCIPDCT 940603.



\newpage

\begin{figure}
\caption
{
\label{fig1gs}
Dependence of the double Gamow--Teller matrix element
$M_{GT}^{2\nu}$ on the renormalization factor in the particle--particle
channel $g_{pp}$. Two different cases for $g_{pp}$ = 0.8 and 1.0 
and for three different QRPA approaches are shown.
Magnitude of the experimental estimate is marked by two parallel
dotted lines \protect{\cite{balysh}}. 
}
\end{figure}

\vskip 2 true cm


\begin{figure}
\caption
{
\label{fig2gs}
Influence of the different number of the added 
multipolarities on $M_{GT}^{2\nu}$ for two types of the
renormalization: RQRPA and SRQRPA.
Full range of the strength factor $g_{pp}$ is scanned
for fixed $g_{ph}$ = 1.0. 
SRQRPA shows less dependence on $g_{pp}$ parameter
than the RQRPA approach.
The experimental values \protect{\cite{balysh}} are between two parallel 
dotted lines.
}
\end{figure}

\vskip 2 true cm


\begin{figure}
\caption
{
\label{fig3gs} 
Contributions of the individual multipolarities 
(1$^+$, 0$^+$, ... , 11$^+$) to the double Gamow--Teller
matrix element within two different RQRPA approaches. 
Calculations were made for $g_{ph}$ =1.0 and for two different
values of the renormalization constant in the particle--particle channel:
$g_{pp}$ = 0.8, 1.0. 
For details see the text.
}
\end{figure}


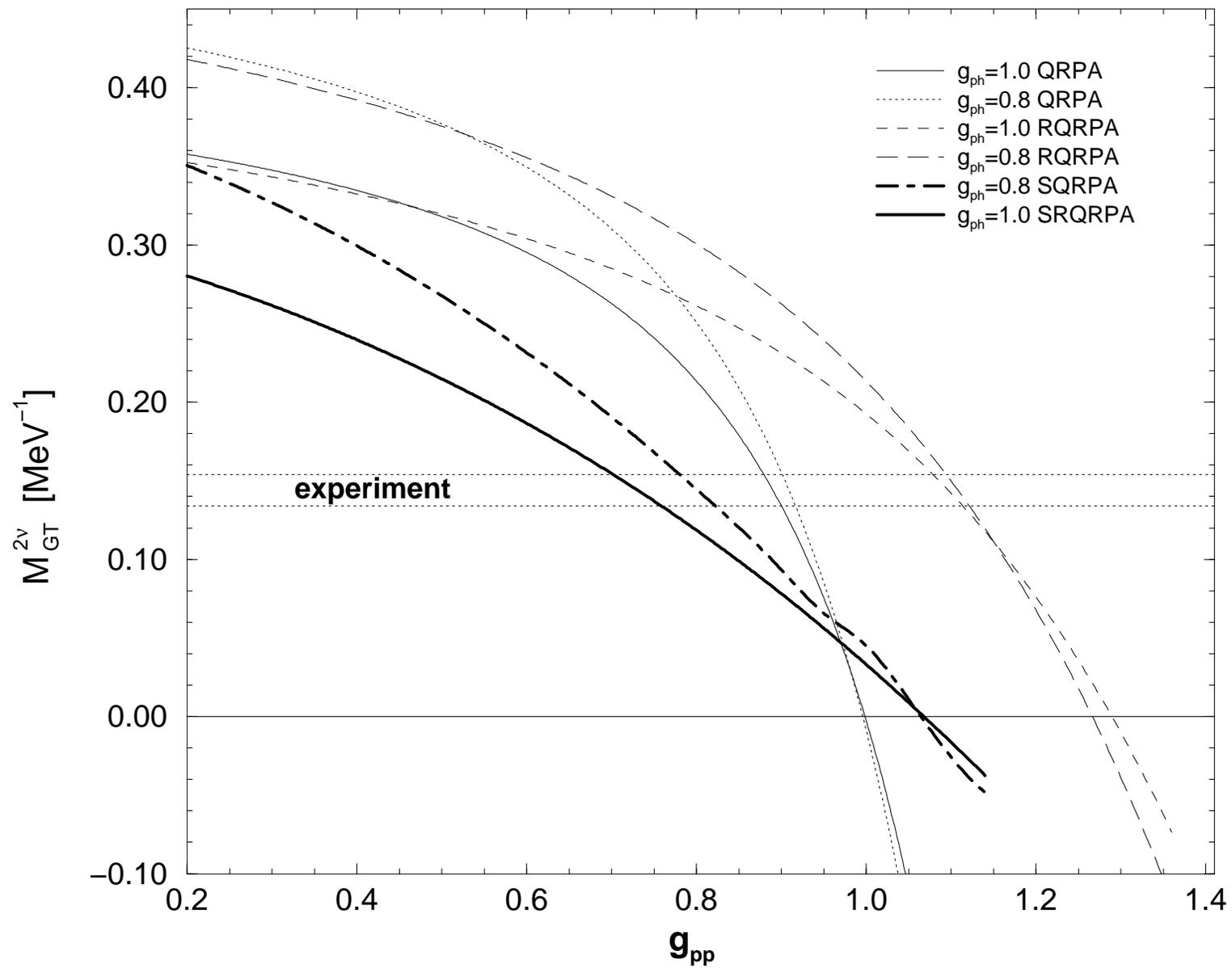
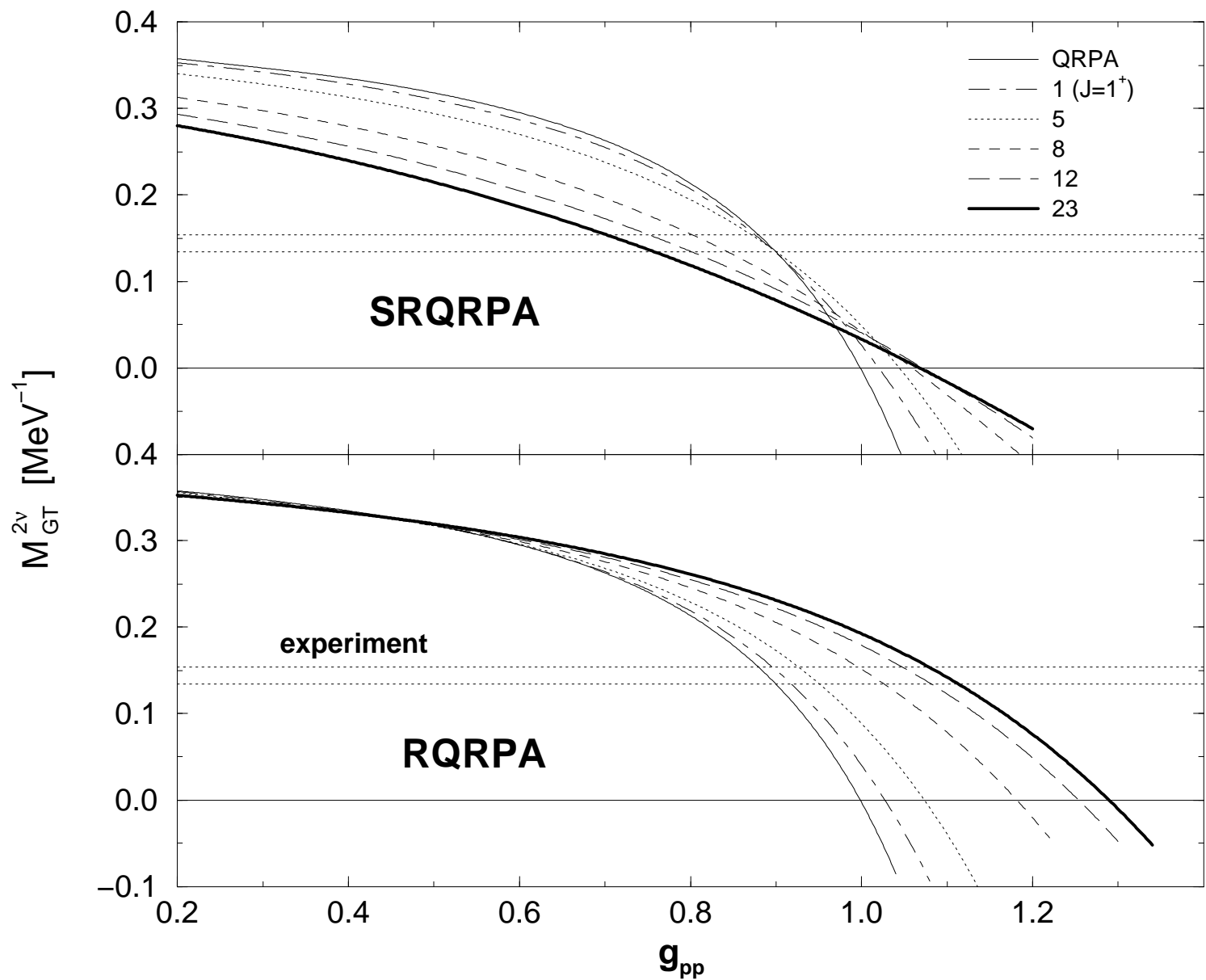
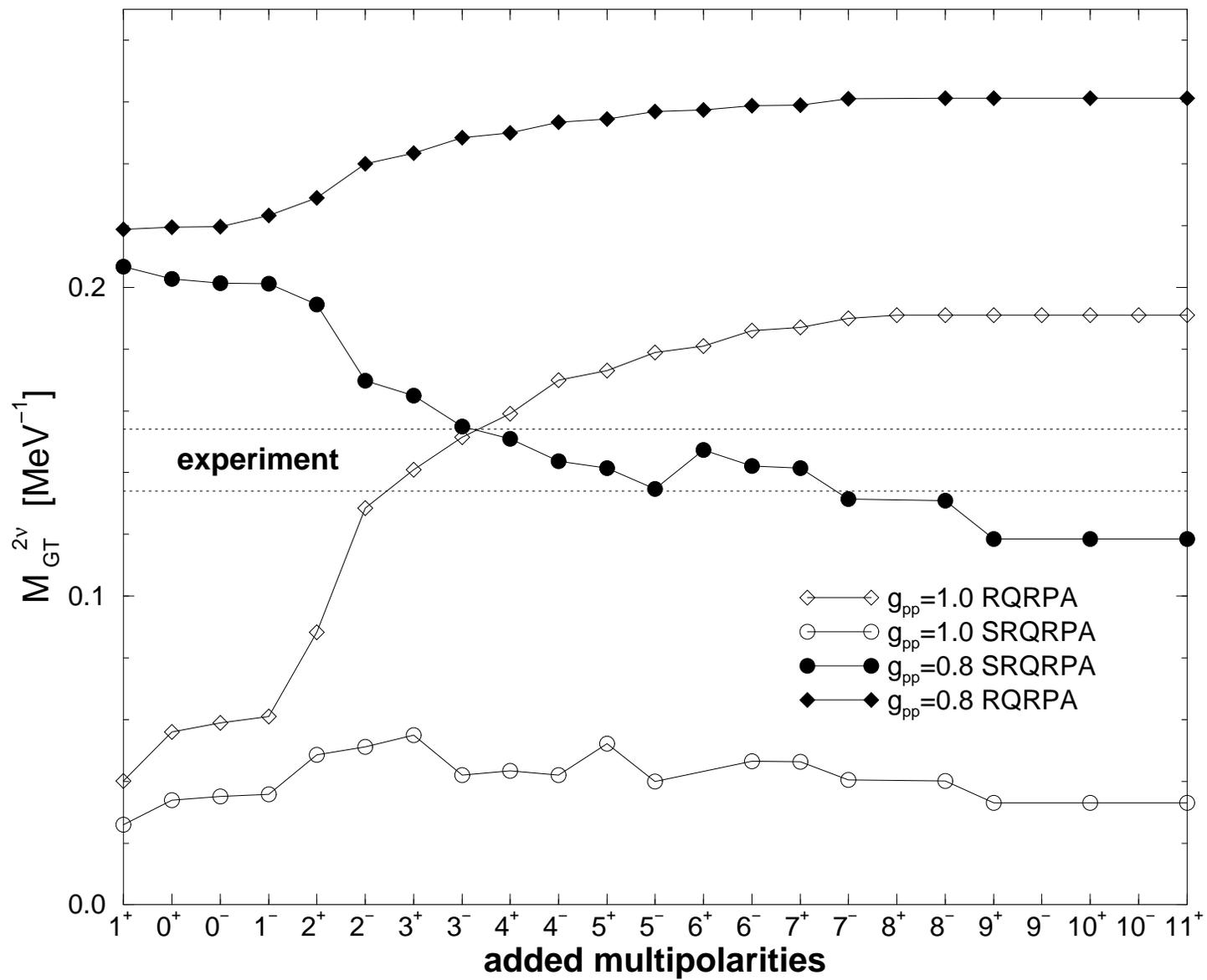
\end{document}